\documentclass[onecolumn,article,11pt]{revtex4}
\usepackage{amssymb}
\usepackage{color}
\usepackage{bm}
\usepackage{graphics,epsfig}
\usepackage{amsmath}
\usepackage{amsfonts,amssymb}
\usepackage{setspace}
\begin{document}

\title{Spinless Quantum Field Theory and Interpretation}
\author{Dong-Sheng Wang}
\email{dongshwa@ucalgary.ca}
\affiliation{Institute for Quantum Science and Technology, Department of Physics and Astronomy,\\ University of Calgary, Alberta T2N 1N4, Canada}

\begin{abstract}
Quantum field theory is mostly known as the most advanced and well-developed theory in physics, which combines quantum mechanics and special relativity consistently.
In this work, we study the spinless quantum field theory, namely the Klein-Gordon equation, and we find that there exists a Dirac form of this equation which predicts the existence of spinless fermion.
For its understanding, we start from the interpretation of quantum field based on the concept of quantum scope, we also extract new meanings of wave-particle duality and quantum statistics.
The existence of spinless fermion is consistent with spin-statistics theorem and also supersymmetry, and it leads to several new kinds of interactions among elementary particles.
Our work contributes to the study of spinless quantum field theory and could have implications for the case of higher spin.
\end{abstract}


\maketitle

\begin{spacing}{1.2}
\section{Introduction}
\label{intro}

Interpretations arise when certain mathematical framework can be endowed with different physics, or certain physical phenomenon can be explained by different mathematics. Lacking of more fundamental principles or just particular mathematical techniques can also lead to misunderstandings and interpretations.
Quantum theory, specifically including quantum mechanics (QM) and quantum field theory (QFT), is always bothered by interpretations almost from its beginning~\cite{Jammer74}.
Some problems are common for QM and QFT, such as the essence of wave-particle duality, the physical meaning of wavefunction (or wavefunctional) etc.
Some problems are due to the issues studied in its particular regime, for the interpretations of QFT, for instance, the meaning of quantum field, and issues of anti-matter.
More fundamental issues are also investigated, such as, whether the foundation of quantum theory lies in QM or QFT, and what the relationship between quantum theory and relativity really is.

In this work we present an interpretation of QFT based on the concept of {\em quantum scope}~\cite{Wang11}.
First I need to explain what basically this concept means, and what is the standpoint to understand quantum theory in this work.
It is reasonable that every kinds of theories are descriptions of motion which capture different aspects of the properties of objects in nature.
Quantum theory is a kind of description which is not the same with classical mechanics.
Scope, which can be understood as wavefunction in the standard language of quantum mechanics, is actually a new kind of universal description of motion.
Roughly speaking, scope of one object captures the {\em ability or propensity} of the motion of the object, and specifies the structure of the motion of the object.
The object, the structure of the object, the motion of the object, and the structure of the motion of the object are connected together yet different in nature, as the result, they need to be described in different ways in physics.
The notion of object can refer to a particle, a collection of particles, or a continuous field.
When we say ``particle'', it can be a point-like particle such as electron, or a package of energy such as photon.
A particle is specified by its mass, charge, shape, spin, and other types of inner degrees of freedom.
The motion of a particle is specified by its coordinate, momentum, and Hamiltonian, and thus a trajectory in spacetime.
The structure or ability of the motion of the particle is just specified by its scope, which is relatively seldom discussed and less understood.
It is worthy to note that the concept of scope says nothing about the particle itself, e.g. whether it is point-like or a string, yet this does not mean the scope can not be affected by the particle itself.

One motivation of our interpretation comes from the relation between quantum theory and relativity.
Usually, QM is viewed as the non-relativistic limit of QFT, for example, the Schr\"{o}dinger equation is viewed as the approximation of Klein-Gordon equation.
This probably leads to the confusion that which equation is the most primary equation in quantum theory.
From the standpoint of scope, we put QM and QFT on the same level, specifically, we treat the generalized form of Schr\"{o}dinger equation as {\em scope equation}.
We show that a new interpretation of the wave-particle duality is possible, particularly, we argue that there are two levels of dualities, propensity-reality duality and scope-particle duality, both of which have explicit physical implications.
The related finding is that the scope equation is precisely equivalent to Heisenberg equation, following from which one essential consequence is that there exist two types of statistics, bosonic and fermionic.
We also analyze the meaning of quantum field, and the differences between field and scope, in order to stress that scope is a rather new concept in physics which needs to be understood properly.

We apply our interpretation in details to Klein-Gordon equation, for which there are well known confusing issues, such as negative probability~\cite{Greiner}, and lots of efforts are made to understand this equation~\cite{FV58,Kostin87,Wha10}.
Still, according to standard theory it is not exactly so clear why a single relativistic particle behaves as boson field, namely, the current form of Klein-Gordon equation describes the motion of spinless relativistic scale field (which can be viewed as a collection of particles) instead of a particle.
With the concept of scope, and by treating the relativistic effect quantally, we are able to express the Klein-Gordon equation in the Hamiltonian form where the Hamiltonian is hermitian.
We name it as Dirac-form of Klein-Gordon equation, since it could be deduced from Dirac equation which describes the dynamics of spin-half fermion.
A significant consequence is that there should be two types of spinless particles, one is described by the usual Klein-Gordon equation as boson, the other is described by the Dirac-form of Klein-Gordon equation as fermion.
The possible existence of spinless fermion is consistent with the spin-statistics theorem.

This work is organized as follows.
We present the scope interpretation in section~\ref{sec:inter}.
In section~\ref{sec:KG} we analyze the Klein-Gordon equation and derive its Dirac-form which describes spinless fermion.
We also discuss some implications of our results in section~\ref{sec:implications}. Then we conclude in section~\ref{sec:conc}.

\section{Interpretation}
\label{sec:inter}

It is mostly expected that quantum phenomenon is unexpected.
We instead take the anthropic standpoint that any reasonable theory exists for us to understand the reasonable world.
As the result, we expect that quantum phenomenon is not weird at all.
In this section, we review and further develop the concept of scope which is originally proposed in Ref.~\cite{Wang11}, and explain our understandings mainly of wave-particle duality, quantum field, the difference between field and scope, also relationship between first quantization and second quantization based on scope.
The formulas apply for general kind of object in nature, for the sake of conciseness, we will just mention single point-like particle in terminology.

The motion of the particle is represented by Hamiltonian $H$.
The scope of the motion of the particle is represented by scope $\Psi$, which is usually called wavefunction or state vector.
The scope equation is
\begin{align}\label{eq:scope}
i\hbar\dot{\Psi}=\hat{H}\Psi,
\end{align}
which is usually known as the general form of Schr\"{o}dinger equation.
Note the standard Schr\"{o}dinger equation simply describes the mechanical motion of non-relativistic particle in one external potential.
The Hamiltonian becomes one operator $\hat{H}$ to generate the spectrum, i.e. states, of the motion.
The scope can be expressed as superposition of all the possible states of the motion.
The term $i\hbar$ ensures that the effect of Hamiltonian will be the phase changes among the states (along with other possible effects such as population exchanges).
Mathematically, the concept of scope contributes trivially to standard quantum theory, yet the interpretational point is that {\em a priori} the scope equation has nothing to do with wave mechanics.

By comparison with Newton equation, which for a single particle takes the form $\vec{F}=m\vec{a}$, the Hamiltonian $\hat{H}$ acts as the external force $\vec{F}$ which generates the evolution of scope $\Psi$, although the scope equation is in first-order derivative of time.
This fact also indicates that $\dot{\Psi}$ does not act as momentum when taking scope $\Psi$ as the coordinate quadrature, which will be clear in the study below.
The scope equation represents a new kind of way, i.e. the ``quantum'' way to study objects, scope captures the structure of the motion of the object which provides all the possibilities or propensity of the object.
This will be much more clear from the study of wave-particle duality.

\subsection{Wave-particle duality}
\label{sec:waveparticle}

The wave-particle duality follows directly from the scope equation.
In literatures and textbooks, there are various understandings in different situations~\cite{Jammer74}, yet the basic notion is that a quantum particle can be both wave and particle simultaneously, and sometimes we can only detect one aspect of the particle sacrificing the other, or, in other words, the motion of a particle has both the particulate and undulatory (wave-like) properties.
To understand this duality, the first point which might be confusing is that wave is strictly not a kind of matter, instead, it is the configuration of the motion of a collection of particles, or it is the way in which energy propagates.
As the result, the duality should be called ``field-particle'', or ``a collection of particles vs. particle'' duality.
However, the puzzle is how can a single particle behaves as a collection of particles?
Should it be ``classical field vs. particle'' or ``quantum field vs. particle''?
Employing our terminology, we interpret the wave-particle duality as {\em scope-matter} duality. Actually, there exist two levels of duality as briefly explained below.

(i) {\em Propensity-Reality duality.}
The motion of the particle is specified by Hamiltonian $H$, which can generate real trajectory in spacetime.
Yet, the ``quantum question'' we can ask is what is the possible ability of $H$?
The propensity of the motion of the particle is specified by scope $\Psi$.
The scope equation~(\ref{eq:scope}) just specifies the connection between reality (Hamiltonian) and propensity (scope).
For this duality, the central mathematical issue is how to quantize the Hamiltonian $H$ to be $\hat{H}$, for which, unfortunately, there still does not exist one standard procedure. In this work we do not involve the quantization problem.
Another issue is to choose the initial value and boundary condition for the scope, which is hard since it is not possible to decide the propensity of a particle merely based on its reality, e.g. one spacetime trajectory. The common choice for initial value is plane wave or Gaussian wave packet.

(ii) {\em Scope-Particle duality.}
The scope itself can also be treated as a ``pseudo'' particle.
The Hamiltonian for the scope is $\mathbb{H}=\Psi^{\dagger}\hat{H}\Psi$, which satisfies Heisenberg equation
\begin{align}\label{eq:heisenbergscope}
i\hbar\dot{\Psi}=[\Psi,\mathbb{H}].
\end{align}
The Hamiltonian $\mathbb{H}$ realizes the scope, and there could be different numbers of the corresponding particles in reality.
The amazing point is that the above equation can be satisfied by two sets of conditions:
\begin{itemize}
\item $[\Psi, \Psi^{\dagger}]=1, [\Psi, \dot{\Psi}]=0$, which is for {\em boson};
\item $[\Psi, \Psi^{\dagger}]_+=1, [\Psi, \dot{\Psi}]_+=0$, which is for {\em fermion}.
\end{itemize}
$[\cdot,\cdot]_{\pm}$ represents commutation (-) and anti-commutation relations (+) respectively, and we often omit the minus sign for commutator.
As the result, the existence of two types of particles are natural.
This duality can be understood in the spirit of usual second quantization, however, here we do not specify any spacetime background, i.e. it is background-independent.
It is clear that $\Psi$, $\Psi^{\dagger}$, and $\dot{\Psi}$ play different roles.
$\Psi^{\dagger}$/$\Psi$ is attributed as the creation/annihilation of a particle.
However, there is no {\em a priori} principle that a particle is equivalent to the scope of the motion of the particle.
As the result, the scope-particle duality leads to the following conjecture:
\newtheorem{Conjecture}{Conjecture}
\begin{Conjecture}
Creation/annihilation of a particle is equivalent to creation/annihilation of the scope of the motion of the particle.
\label{con:1}
\end{Conjecture}
The above conjecture is not clearly stated in standard quantum theory, which might blur the physical meanings of wavefunction also first and second quantization.

From the scope-particle duality, Schr\"{o}dinger equation is equivalent to Heisenberg equation.
This is different from the problem that whether the Schr\"{o}dinger picture and Heisenberg picture to deal with time evolution are equivalent or not, since the notion of observable is needed.
It is direct to understand the equivalence of Schr\"{o}dinger picture and Heisenberg picture since the observation effects in different pictures are equivalent, no matter which picture is more primary.
The equivalence of Schr\"{o}dinger equation and Heisenberg equation, which is not stressed before in literatures and textbooks, is closely related to statistics yet not to spin, and also highlights the fundamental roles of non-hermitian operators.

\subsection{Hamiltonian and Lagrangian forms}
\label{sec:hamlag}

Since QFT is often expressed in the language of Lagrangian, here we present that for scope equation.
Hamiltonian and Lagrangian forms of mechanics are universal, that is, they are not restricted to be classical, quantum, or relativistic.
The underlying principle is the {\em stationary of action}.

The Lagrangian for scope equation is
\begin{align}\label{eq:lscope}
\mathbb{L}=\Psi^{\dagger}(i\hbar\partial_t-\hat{H})\Psi.
\end{align}
It is direct to check that Euler-Lagrangian equation is equivalent to scope equation.
Define $\Pi=\frac{\partial \mathbb{L}}{\partial \dot{\Psi}}=i\hbar\Psi^{\dagger}$, then $\mathbb{H}=\Pi\dot{\Psi}-\mathbb{L}=\Psi^{\dagger}\hat{H}\Psi$, as the result,
\begin{align}\label{eq:hamiquantum}
\frac{\partial \mathbb{H}}{\partial \Psi}=-\dot{\Pi}, \;
\frac{\partial \mathbb{H}}{\partial \Pi} =\dot{\Psi},
\end{align}
which is the Hamiltonian equation, also equivalent to scope equation.
Although the Hamiltonian and Lagrangian forms are very similar with those for a classical field, the crucial difference is that the momentum quadrature $\Pi$ is specified by the conjugate of scope $\Psi$ instead of its time derivative $\dot{\Psi}$.
This manifests that scope is different with field, which is further discussed later on.

\subsection{Scope under representation}
\label{sec:scoprep}

The quantization from $H$ to $\hat{H}$ mentioned above brings that quantum observables are operators.
The existence of representation in quantum theory relies on the fact that observables can be non-commutable, which means some observables can not be measured simultaneously, such as position and momentum for a particle.
One way to manifest the non-commutability of operators is the uncertainty principle.
The scope $\Psi$ and related momentum $\Psi^{\dagger}$ does not commute both for the bosonic and fermionic cases.
We can easily show that the vacuum is the minimally uncertain state.
Let the standard deviation be $\Delta\Psi=\Psi-\langle\Psi\rangle$, $\Delta\Psi^{\dagger}=\Psi^{\dagger}-\langle\Psi^{\dagger}\rangle$, with $\langle\cdot\rangle=\langle\Omega|\cdot|\Omega\rangle$, $|\Omega\rangle$ is vacuum state.
From $\Psi|\Omega\rangle=0$, $\langle\Omega|\Psi^{\dagger}=0$, we can find $\Delta\Psi\Delta\Psi^{\dagger}=1/2$.
The general case  $\Delta\Psi\Delta\Psi^{\dagger}>1/2$ can be directly proved employing Cauchy-Schwarz inequality.

The scope equation~(\ref{eq:scope}) does not specify any representation.
Any solvable scope equation in practice needs to be equipped with some representations, including spin, momentum, angular momentum, and energy (occupation number), etc.
In quantum theory, the coordinate forms the background for the scope as well as the dynamical variable of the particle.
For any certain time, the coordinate forms a complete set of basis as
\begin{align}\label{eq:xbasis}
\int d^3x \;|x\rangle\langle x|=I,
\end{align}
$I$ represents identity which could be finite or infinite dimensional.
In Dirac notation, write scope as $|\Psi\rangle$, and then $|\Psi\rangle=\int d^3x |x\rangle\langle x|\Psi\rangle=\int d^3x \Psi(x) |x\rangle$, $\Psi(x)\equiv\langle x|\Psi\rangle$ is just the usually called wavefunction or quantum field.
Also, there will be a Lagrangian density $\mathcal{L}=\Psi^{\dagger}(x)(i\hbar\partial_t-\hat{H}(x))\Psi(x)$ such that $\mathbb{L}=\int d^3x \mathcal{L}$, and a Hamiltonian density $\mathcal{H}=\Psi^{\dagger}(x)\hat{H}(x)\Psi(x)$ such that $\mathbb{H}=\int d^3x\; \mathcal{H}$.

$|\Psi(x)|^2d^3x$ is interpreted as the weight (probability) of a particle to exist (created or annihilated in second quantization, move mechanically in first quantization) in regime $d^3x$.
The reason for various interpretations are mainly that scope describes the ability of the motion of a particle, regardless of the detailed styles of movements.
In other words, the description based on scope is so general that it can describe any kind of movement in principle.

Scope is constrained by normalization condition, which leads to the statistical interpretation.
In condensed-matter physics, the scope is normalized to the number of particles instead of unity for the probability.
The reason is that each particle, e.g. electron, is treated as identical particle and thus each particle has the same scope, as the result, the single-particle scope is employed to describe the collection of identical particles except that the normalization is not unity.
The single-particle scope can potentially create or annihilate the particle in different states, and the collection of particles is treated as the collection of the realized states of a single particle.
The interactions among the particles are treated as self-interaction of the scope, contributing nonlinear terms in the scope equation.
However, as we know, the notion of identity particle is actually a postulate in quantum theory, in reality particles like electrons might be different from each other in the sense that a single-particle scope cannot be employed to describe collection of particles.
The above argument is stated as the following conjecture relating to statistics.
\begin{Conjecture}
For a collection of identical particles each with scopes $\Psi(x_i)$, $i=1,\dots,n$, the scope for the whole system is $\Psi(x_1,x_2,\dots,x_n)$, such that it is symmetric for boson and anti-symmetric for fermion with respect to exchange of any two particles.
\label{con:2}
\end{Conjecture}
One might find the analysis above does not make significant difference with standard quantum theory mathematically, the point here is we try to explain the quantum description based on the view of scope.
The above conjecture is consistent with the propensity-reality duality, in the sense that the scope for the whole system specifies its propensity, and the reality is specified by a corresponding Hamiltonian which includes the self-interactions among the particles.
It is also consistent with the scope-particle duality, as we stated above, that all the particles are treated as the collection of the realized states of a single particle.

\subsection{Scope and quantum field}
\label{sec:scopfield}

For the notion of quantum field, confusions always exist, such as in Refs.~\cite{Bak09,Hob12}, here we develop our understandings based on scope.
There is one non-trivial difference between a field and a scope.
As we know, a classical field $\phi(x)$ is a continuous collection of particles $q_i$.
The letter $\phi$ represents the coordinate of the underlying particles, and letter $x$ just substitutes the index $i$.
The momentum of the field is usually $\dot{\phi}(x)$, similar with that for a single particle as $\dot{q}_i$.
A field $\phi(x)$ can be quantized as $\hat{\phi}(x)$, which can act on a scope for the field.
However, a scope is totally different as we discussed above.
The scope $\Psi$ cannot be quantized since there cannot be a ``scope of scope'' (note quantization means the introduction of scope).
The so-called quantum field $\Psi(x)$ is actually the scope in the basis of $|x\rangle$.
The corresponding momentum relates to $\Psi^{\dagger}(x)$ instead of $\dot{\Psi}(x)$.
The quantum field is counterfactual which potentially exists and specifies the possible states of the particle, according to the propensity-reality duality.
Yet, at the same time, according to the scope-particle duality, a scope can also be associated with a Hamiltonian $\mathbb{H}$ which can turn scope from propensity into reality. That is to say, a scope can be potentially exist or has been realized in different practical situations.
As the result, in the usual textbook terminology, there exist actually two types of quantum fields: one is the quantized classical field, such as a scale field, electromagnetic field, the other is scope under coordinate representation, which is automatically quantum, and similar for their Fourier transforms.
Furthermore, it turns out the quantized classical field is boson, and scope is fermion, which is consistent with the fact that usually bosonic particle is the messenger which propagates interactions among fermionic particles.

\section{Klein-Gordon equation}
\label{sec:KG}

In the following we apply our interpretation to Klein-Gordon equation.
Before that, we need to treat relativistic effect quantally in order to express Klein-Gordon field equation in the form of scope equation.
Then we demonstrate that the Klein-Gordon scope has to be fermion.

\subsection{Spacetime qubit}
\label{sec:sptimqubi}

Special relativity of a single particle reveals there is a fundamental limitation of its ``freedom'', that is to say, there exists a kind of internal coherence which depends on the velocity of the particle.
The concept of velocity is based on time and (space) coordinate.
To incorporate special relativity into quantum theory, we need not to treat special relativity as a generalization of classical mechanics.
That is to say, we need to interpret special relativity based on the description of scope.
It turns out the effect of special relativity can be treated quantally, the spacetime state can be expressed as one two-level system, i.e. qubit.
Define the spacetime qubit as
\begin{align}\label{eq:sptqubit}
\hat{s}=\begin{pmatrix}
  ct+z & x-iy \\ x+iy & ct-z
\end{pmatrix}=ct\sigma_0+x\sigma_x+y\sigma_y+z\sigma_z,
\end{align}
$\sigma_0=I$, $\sigma_{x,y,z}$ are Pauli matrices. The norm of $\hat{s}$ is the proper distance $\|s\|=c^2t^2-r^2$.
The trace is $\text{tr}\hat{s}=2ct$, a normalized spacetime qubit can be defined as $\hat{s}/2ct$.
Note the above formula is also studied in other contexts with different motivations~\cite{Kim12}, and different notions of spacetime qubit also exist in literatures~\cite{PMR11}.
The qubit can be labeled on the Bloch ball, as the result, light can be viewed as a pure spacetime qubit forming the sphere, and the particle will fall into the ball, the center of the ball is for particle at rest.
The Lorentz transformation is a ``determinant-preserving'' channel, neither trace preserving, nor completely positive channel.
In the basis of $\sigma_0$ and Pauli matrices, the spacetime qubit can be represented as a column vector $\vec{s}=(ct,x,y,z)$, and then Lorentz channel will take the form of so-called $\mathbb{T}$ matrix~\cite{King2001}, which coincides with the standard form of Lorentz transformation~\cite{Greiner}.

A mass qubit can also be defined as
\begin{align}\label{eq:sptqubit}
\hat{m}=\begin{pmatrix}
  E-p_zc & p_xc-ip_yc \\ p_xc+ip_yc & E+p_zc
\end{pmatrix}.
\end{align}
The norm of $\hat{m}$ is the rest energy $\|s\|=m^2c^4$.

Usually, relativistic quantum mechanics is treated as the extension of standard quantum mechanics.
However, by the concept of spacetime qubit, the effect of relativity will be represented by a two-dimensional Hilbert space.
In the following, we will show that there exists the basis of $|\circ\rangle$ (anti-particle) and $|\bullet\rangle$ (particle) for the Klein-Gordon Hamiltonian.

\subsection{Dirac-form of Klein-Gordon equation}
\label{sec:DKG}

The Klein-Gordon equation is usually interpreted as a classical relativistic massive scale field equation, which takes the form
\begin{align}\label{eq:kg}
(\Box+m^2c^2/\hbar^2)\phi(x)=0,
\end{align}
with d'Alembertian operator $\Box=\frac{1}{c^2}\frac{\partial^2}{\partial t^2}-\nabla^2$.
The Klein-Gordon equation is derived from Einstein's relation $E^2=p^2c^2+m^2c^4$, by $E\rightarrow i\hbar\frac{\partial}{\partial t}$, $p\rightarrow i\hbar \nabla$.
This equation can be viewed as a generalization of a free scale field equation from massless to massive case, or a generalization of Schr\"{o}dinger equation to relativistic case, and also can be treated as a generalization of a classical field in a certain medium to electromagnetic case (with $\hbar$ and $c$).
It is often accepted that the Klein-Gordon field $\phi(x)$ is classical since it is not one operator yet.
However there is $\hbar$ in the equation~(\ref{eq:kg}) which means there exists some quantum essence.
This apparent conflict can be resolved when we convert the Klein-Gordon equation into a scope equation under the representation of $|\bullet,\circ\rangle$.
Next we derive the new formula.

Let $\phi(x)=\psi_1(x)+\psi_2(x)$, and each of the two satisfies Klein-Gordon equation~(\ref{eq:kg}).
The Klein-Gordon scope is defined in the basis of $|x\rangle$ and $|\bullet,\circ\rangle$ as
\begin{align}\label{eq:kgscope}
|\Psi\rangle_{\text{KG}}=\psi_1(x)|\bullet\rangle+\psi_2(x)|\circ\rangle,
\end{align}
or $|\Psi\rangle_{\text{KG}}=\begin{pmatrix} \psi_1(x) \\ \psi_2(x) \end{pmatrix}$, the up branch is for particle, the down for anti-particle.
Suppose
\begin{equation}
\left\{ \begin{array}{lll}
i\hbar\dot{\psi}_1(x)&=&a\psi_1(x)+b\psi_2(x),  \\ i\hbar\dot{\psi}_2(x)&=&c\psi_2(x)+d\psi_1(x),
\end{array} \right.
\end{equation}
$a, b, c, d$ are parameters to be determined.
The above coupled equation is equivalent to scope equation with Hamiltonian $\hat{H}=\begin{pmatrix} a & b \\ d & c \end{pmatrix}$.
The four parameters need to satisfy two conditions:
i) Klein-Gordon equation~(\ref{eq:kg});
ii) eigenvalues of $\hat{H}$ are $E_{\pm}=\pm\sqrt{p^2c^2+m^2c^4}=\left(a+c\pm\sqrt{(a-c)^2+4bd}\right)/2$.
We find $c^2-a^2=(a+c)(d-b)$. There are several possible solutions.
\begin{itemize}
\item $a=c$, we find $\hat{H}=\begin{pmatrix} 0 & E \\ E & 0 \end{pmatrix}$, with $E=\sqrt{p^2c^2+m^2c^4}$.
    Diagonalize as $\hat{H}_d=E\sigma_z$.
    This is well known as the relativistic Schr\"{o}dinger form.
    The time derivation is $\dot{\phi}(x)=\frac{E}{i\hbar}\phi(x)$, and we can find the solution $\phi(x,t)=e^{-iEt/\hbar}\phi(x,0)$.
    In this case, the roles of $\psi_1(x)$ and $\psi_2(x)$ are trivial.
\item $a=-c$$\Rightarrow$$E_{\pm}=\pm\sqrt{a^2+bd}$.
\begin{itemize}
\item $a=mc^2+p^2/2m$, $b=-d=p^2/2m$ $\Rightarrow$ $\hat{H}=\begin{pmatrix} mc^2+p^2/2m & p^2/2m \\ -p^2/2m & -mc^2-p^2/2m \end{pmatrix}$.
      This is usually called the Hamiltonian form, and the diagonalized form is called Feshbch-Villars representation~\cite{FV58}.
      This form is not good enough since the mass $m$ cannot be reduced to zero for the case of light, and the Hamiltonian is not hermitian.
      The time derivation is $\dot{\phi}(x)=\frac{mc^2}{i\hbar}(\psi_1(x)-\psi_2(x))$.
\item $a=mc^2$, $b=d=|p|c$ $\Rightarrow$ $\hat{H}=\begin{pmatrix} mc^2 & |p|c \\ |p|c & -mc^2 \end{pmatrix}$, $|p|=\sqrt{p_x^2+p_y^2+p_z^2}$.
    This can be named as ``Dirac form'' since it can be derived from Dirac equation.
    The time derivation is $\dot{\phi}(x)=\frac{mc^2}{i\hbar}(\psi_1(x)-\psi_2(x))+\frac{|p|c}{i\hbar}(\psi_1(x)+\psi_2(x))$.
\end{itemize}
\end{itemize}
The reason for there could be different forms of Hamiltonian is that both values of $\phi(x)$ and $\dot{\phi}(x)$ are required to solve the Klein-Gordon equation, we see that the above forms yield different $\dot{\phi}(x)$.
The Dirac form can be easily reduced to the massless case for light and speed-zero case for a rest particle, as the result, it is preferred.
Thus, the Klein-Gordon equation can be written in Dirac form as
\begin{align}\label{eq:kgd}
i\hbar|\dot{\Psi}\rangle_{\text{KGD}}=\begin{pmatrix} mc^2 & |p|c \\ |p|c & -mc^2 \end{pmatrix}|\Psi\rangle_{\text{KGD}}.
\end{align}
We use $\mathrm{KGD}$ to denote ``Dirac-form of Klein-Gordon''.
There are several obvious consequences.
One is that for a rest particle, the eigenstates are anti-particle $|\circ\rangle$ with negative rest energy and particle $|\bullet\rangle$ with positive rest energy.
For massless light, the eigenstates are superpositions of $|\bullet,\circ\rangle$ as $|\pm\rangle=\frac{\sqrt{2}}{2}(|\circ\rangle\pm|\bullet\rangle)$, that is, light can be viewed as a maximally superposed states of particle and anti-particle (regardless of the polarization i.e. spin of the light).
When a particle acquires higher and higher velocity, it will eventually becomes a superposed state of particle and anti-particle, i.e. a flash.

Klein-Gordon scope equation~(\ref{eq:kgd}) can be derived from Dirac equation.
Dirac scope is under the representation of $|x\rangle$, $|\bullet,\circ\rangle$, and spin $|\uparrow,\downarrow\rangle$ as
\begin{align}\label{eq:scopedirac}
|\Psi\rangle_{\text{D}}=\varphi_1(x)|\bullet,\uparrow\rangle+\varphi_2(x)|\bullet,\downarrow\rangle+\varphi_3(x)|\circ,\uparrow\rangle+\varphi_4(x)|\circ,\downarrow\rangle,
\end{align}
Dirac equation describes the motion of a massive charged spin-half particle, e.g. electron, it takes the form
\begin{align}\label{eq:dirac}
i\hbar|\dot{\Psi}\rangle_{\text{D}}=(\vec{\alpha} \cdot \vec{p}c+\beta mc^2)|\Psi\rangle_{\text{D}},
\end{align}
with Hamiltonian $H_{\text{D}}=\vec{\alpha} \cdot \vec{p}c+\beta mc^2$ and
\begin{align}\label{eq:dirachar}
H_{\text{D}}=\begin{pmatrix} mc^2 & 0 & cp_z & c(p_x-ip_y) \\ 0 & mc^2 & c(p_x+ip_y) & -cp_z \\ cp_z & c(p_x-ip_y) & -mc^2 & 0 \\ c(p_x+ip_y) & -cp_z & 0 & -mc^2\end{pmatrix},
\end{align}
Dirac matrices are
\begin{align}\label{eq:diracma}\nonumber
\alpha^0&\equiv\beta\equiv\gamma^0=\sigma_z\otimes I=\begin{pmatrix} I & 0 \\ 0 & -I\end{pmatrix},\\ \nonumber
\alpha^1&=\sigma_x\otimes \sigma_x=\begin{pmatrix} 0 & \sigma_x \\ \sigma_x & 0\end{pmatrix},\alpha^2=\sigma_x\otimes \sigma_y=\begin{pmatrix} 0 & \sigma_y \\ \sigma_y & 0\end{pmatrix},\\  \alpha^3&=\sigma_x\otimes \sigma_z=\begin{pmatrix} 0 & \sigma_z \\ \sigma_z & 0\end{pmatrix}.
\end{align}
The Klein-Gordon scope equation~(\ref{eq:kgd}) can be derived by the following trick: define operator $W=(p_x\sigma_x+p_y\sigma_y+p_z\sigma_z)/|p|$, then $W^2=I$, $H_{\text{D}}$ can be reduced as $H_{\text{D}}=(mc^2\sigma_z+|p|c\sigma_x)\otimes I$, the identity represents the information of spin which is then ignored, and the rest part is just the Klein-Gordon Hamiltonian.
The above procedure corresponds to the reduction of Lorentz group from SO(1,3) to SO(1,1).
The Dirac-form of Klein-Gordon equation can also be understood as 1-dimensional Dirac equation mathematically, however, physically the difference is a Klein-Gordon fermion, demonstrated below, takes momentum as scale and couples to other fields in new ways.

\subsection{Spinless fermion}
\label{sec:spinlessfermion}

As we know, the Klein-Gordon field is boson, for charged case, the Hamiltonian is $\mathbb{H}_{\text{KG}}=\int d^3x [\dot{\phi}(x)\dot{\phi}^{\dagger}(x)+\nabla\phi^{\dagger}(x)\nabla\phi(x)+m^2c^2\phi(x)\phi^{\dagger}(x)/\hbar^2]/2$, the quantized field takes the form
\begin{align}\label{eq:kgfield}
\hat{\phi}(x)=\int\frac{d^3p}{(2\pi)^3}\frac{1}{\sqrt{2E}} \left\{\hat{a}_p e^{-i(px+Et)/\hbar}+ \hat{c}_p^{\dagger} e^{i(px+Et)/\hbar}\right\},
\end{align}
with $E=\sqrt{m^2c^4+p^2c^2}$.
The Hamiltonian reduces to the form
\begin{align}\label{eq:kghami}
\mathbb{H}_{\text{KG}}=\int\frac{d^3p}{(2\pi)^3}\frac{E}{2}\left(\hat{a}_p^{\dagger}\hat{a}_p+\hat{c}_p^{\dagger}\hat{c}_p+1\right),
\end{align}
with commutation relation $[\hat{a}_p,\hat{a}_{p'}^{\dagger}]=\delta^3(p-p')$, $[\hat{c}_p,\hat{c}_{p'}^{\dagger}]=\delta^3(p-p')$.
The rather hidden problem is that why the issue of negative energy does not show up.
Recall that the field $\phi(x)$ is the sum of two terms, $\psi_1(x)$ for particle, and $\psi_2(x)$ for anti-particle.
$\psi_1(x)$ corresponds to $\hat{a}_p$, and $\psi_2(x)$ corresponds to $\hat{c}_p^{\dagger}$, that is, the expansion~(\ref{eq:kghami}) has employed the assumption that a creation of anti-particle is equivalent to the annihilation of particle.
As the result, all particles can be viewed as the same kind and then there is no problem with negative energy.
However, a non-trivial issue is that a creation of anti-particle corresponds to, yet is not equivalent to, the annihilation of particle, which will then leads to the coexistence of particle and anti-particle, and also the spinless fermion.

It is natural that the Klein-Gordon scope should be fermion, similar with Dirac fermion.
Quantize the Klein-Gordon scope as
\begin{align}\label{eq:kgfield}
\Psi_{\text{KGD}}=\int\frac{d^3p}{(2\pi)^3}\frac{1}{\sqrt{2E}} \left\{\hat{f}_p |\xi_p\rangle e^{-i(px+Et)/\hbar}+ \hat{g}_p^{\dagger}  |\eta_p\rangle e^{i(px+Et)/\hbar}\right\},
\end{align}
the two eigenstates are $|\xi_p\rangle=\left(\sqrt{mc^2+E},\frac{|p|c}{\sqrt{mc^2+E}}\right)$ for particle, $|\eta_p\rangle=\left(-\sqrt{E-mc^2},\frac{|p|c}{\sqrt{E-mc^2}}\right)$ for anti-particle, the Hamiltonian takes the form
\begin{align}\label{eq:kgdhami}
\mathbb{H}_{\text{KGD}}=\int\frac{d^3p}{(2\pi)^3}E\left(\hat{f}_p^{\dagger}\hat{f}_p+\hat{g}_p^{\dagger}\hat{g}_p-1\right),
\end{align}
with anti-commutation relation $[\hat{f}_p,\hat{f}_{p'}^{\dagger}]_+=\delta^3(p-p')$, $[\hat{g}_p,\hat{g}_{p'}^{\dagger}]_+=\delta^3(p-p')$.
By comparison, the Dirac field~\cite{PS95} can be viewed as generalization of Klein-Gordon scope to include the spin-dependent (up and down) terms. In addition, it is direct to check that the CPT theorem holds for Klein-Gordon fermion.

\section{Implications}
\label{sec:implications}

In this section, we briefly discuss some immediate consequences of the existence of Klein-Gordon fermion.

\subsection{Spin-Statistics theorem}\label{sec:ssthe}

As shown in section~\ref{sec:waveparticle}, there exist two kinds of statistics, without referring to spin.
The spin-statistics theorem states that particle with integral spin is boson, while particle with half-integral spin is fermion.
The simplest and most fundamental proof of this theorem is provided by Weinberg~\cite{MR03} based on {\em microcausality}, without any reference to positivity of energy and Dirac's hole theory.
The microcausality states that measurements (or events) at space-like points can not disturb each other, which can be expressed as $[\psi(x),\psi(x')]_{\pm}=[\psi(x),\psi^{\dagger}(x')]_{\pm}=0$, for $(x-x')^2<0$, consistent with relativity.
It is direct to verify that the Klein-Gordon scale field in equation~(\ref{eq:kg}) has to be boson, and the Dirac scope in equation~(\ref{eq:dirac}) has to be fermion.
Since the Dirac-form of Klein-Gordon scope in equation~(\ref{eq:kgd}) is in the same category with Dirac scope, it must be that the Klein-Gordon scope is fermion.
Consequently, one confusion arises: given a spinless particle, is it boson or fermion?
For a spinless particle, if it is not relativistic, i.e. it is a Schr\"{o}dinger particle, then it can be boson as well as fermion, and actually it satisfies Boltzmann statistics.
The main reason is that the relativistic invariance does not work, or, in our language, there is no spacetime qubit.
For a Klein-Gordon particle, however, the anti-particle state is involved.
There could be three possibilities: i) a Klein-Gordon particle can be fermion or boson in different situations; ii) a Klein-Gordon particle can be fermion and boson at the same time; iii) there are two kinds of Klein-Gordon particles, one is fermion, the other is boson.
The most possible case is the last one.

On the landscape of spin-statistics connection, we need to separate the spinless case from other cases. The spin-statistics theorem still applies for spinless case since the underlying principle of microcausality is universal.

\subsection{Ghosts}\label{sec:ghost}

There exists the so-called Faddeev-Popov ghost which is spinless scale and manifest fermion statistics~\cite{FP67}.
Clearly the Faddeev-Popov ghosts violate the spin-statistics theorem.
There are various kinds of ghosts thus kinds of technical reasons for their existence.
The Faddeev-Popov ghosts arise from the redundancy of field configurations related by gauge symmetries for path-integral formula, the ghosts are introduced to break the gauge symmetry.
The ghosts only appear in loops as virtual particles in Feynman diagrams. 
The Klein-Gordon scope and the Faddeev-Popov ghosts are both spinless fermion.
However, the Klein-Gordon scope is a two-component ``spinor'' instead of a scale, thus the Klein-Gordon fermion is not a ghost, i.e. it is a real stable particle.

Technically, the Klein-Gordon fermion can be viewed as the spinor of two free scale fermionic ghosts.
As the result, the Klein-Gordon fermion might be used to attack problems associated with ghosts by modifying the scale ghosts to the Klein-Gordon fermion.

\subsection{Supersymmetry}\label{sec:ssymmetry}

It turns out our results have some implications on supersymmetry, which mainly states that every particle has a corresponding superpartner with different statistics and the same mass~\cite{Wei99}.
The Klein-Gordon fermion and boson form a pair of superpartners for each other, as the result, it supports supersymmetry in a quite trivial way since there is no spin.
Thus we remain quite neutral for the validity of supersymmetry since more efforts on the scope with spin are necessary.
Notably, when we combine the Hamiltonian~(\ref{eq:kghami}) and (\ref{eq:kgdhami}) together, the infinite zero-point energy cancels out, if the first one is re-scaled by two times, which is equivalent to a re-scaling of the scope~(\ref{eq:kgscope}) by $\frac{\sqrt{2}}{2}$.
This provides a way to deal with the divergence and renormalization problem.

\subsection{Interaction}\label{sec:interaction}

In QFT, interaction is usually specified by the form of Lagrangian $\mathbb{L}$.
Next we construct several new kinds of interactions due to the existence of Klein-Gordon fermion, and equivalent Hamiltonian $\mathbb{H}$ for Hamiltonian mechanics also Heisenberg equation and $\hat{H}$ for scope equation can be constructed too.
The Klein-Gordon fermion has the Lagrangian $\mathbb{L}_{\text{KGD}}=\Psi^{\dagger}(i\hbar\partial_t-|p|c\sigma_x-mc^2\sigma_z)\Psi$.
The first kind of legal interaction is between Klein-Gordon fermion and Klein-Gordon boson as
\begin{align}\label{eq:intkgferbon}
\mathbb{L}=\mathbb{L}_{\text{KGD}}+\mathbb{L}_{\text{KG}}-w\Psi^{\dagger}\sigma_z\Psi\phi,
\end{align}
with $w$ as coupling constant, the Lagrangian for Klein-Gordon boson as $\mathbb{L}_{\text{KG}}=\dot{\phi}\dot{\phi}^{\dagger}+\nabla\phi^{\dagger}\nabla\phi-m^2c^2\phi\phi^{\dagger}/\hbar^2$, has been enlarged twice.
The above interaction is similar with Yukawa interaction~\cite{PS95}, which describes the interaction between a Klein-Gordon boson and Dirac fermion.
A different type of interaction is also possible, if $\sigma_z$ in the last term of equation~(\ref{eq:intkgferbon}) is absent.

When the Klein-Gordon fermion is charged with charge $e$, there could be another kind of interaction which is between the fermion and gauge field.
Different from QED (quantum electrodynamics), here the fact is there is no spin-momentum coupling, and momentum plays the role by its magnitude.
As the result, it is natural to suppose that the Klein-Gordon fermion cannot distinguish the directions of the vector potential $\vec{A}$.
The Lagrangian takes the form
\begin{align}\label{eq:intkgferphoton}
\mathbb{L}=\mathbb{L}_{\text{KGD}}-\frac{1}{4}F^{\mu\nu}F_{\mu\nu}-e\Psi^{\dagger}\Psi\phi-ec\Psi^{\dagger}\sigma_x\Psi A,
\end{align}
$\phi$ is scale potential, $A$ is the magnitude of the vector potential.
Similar with that in QED, the scale and vector potentials play different roles.

In addition, as we mentioned above, the Klein-Gordon fermion can also be employed to study problems involving ghosts.
Detailed studies of these interactions need further analysis.

\section{Discussion and Conclusion}
\label{sec:conc}

In this work we studied the quantum field theory for spinless case, and we found the Dirac-form of Klein-Gordon equation and discussed some implications.
We started from the interpretations based on scope for quantum field theory, since it is crucial for the discovery of Klein-Gordon fermion.
During the development of quantum field theory the Klein-Gordon fermion is absent, there might be several reasons for this.
(i) The physical meaning of wavefunction (and wavefunctional) is not clear, e.g. the difference between wavefunction and field.
(ii) The effects of special relativity are not treated in a quantum way.
(iii) The dynamics in quantum field theory is usually expressed in Lagrangian form, instead of Schr\"{o}dinger equation and a Hamiltonian, and it is believed that the condition of hermiticity needs to be generalized to pseudo-hermiticity.
The Klein-Gordon fermion takes one concise Hamiltonian and has explicit physical essence.
As we have demonstrated, it is consistent with spin-statistics theorem, and actually develops it for spinless case, and it also supports supersymmetry, which is believed to be a fundamental symmetry in nature.

In our interpretation, we have listed separately two conjectures, here we discuss their implications further.
The first conjecture is seldom mentioned in standard theory, yet in practice related problems could be encountered, e.g., we find that quantum theory says nothing about the volume of electron, the shape of single photon etc.
This conjecture indicates the limitation of quantum theory, for instance, quantum theory can not describe how a particle is created or annihilated, how a photon is emitted within a finite lifetime, for which new kind of dynamics is needed.
The second conjecture relates to the identity postulate, which provides a foundation for the application of second quantization in condensed-matter physics.
However, the statement of this conjecture has actually been weakened to a certain degree since different kinds of statistics (such as anyon) exist in practice.
On the macroscopic level, the effects of the difference between particles like electrons could be trivial, as the result, the second conjecture is still valid and does not set a fundamental limitation of quantum theory.

The scope interpretation provides a general framework to understand and employ quantum theory, we have managed to put quantum mechanics and quantum field theory on the same footing and treat relativity quantally in this work.
For the case with spin, our interpretation can also be employed.
It is direct to check that the gauge field (photon) is specified by scope under representation of spin, occupation number, and momentum.
For QCD (quantum chromodynamics) and non-Abelian gauge field, for every interaction there exists one Hamiltonian for scope equation, and then the dynamics can be investigated employing scope equation instead of Lagrangian.

\end{spacing}

\end{document}